\label{key}
\documentclass[conference]{IEEEtran}
\IEEEoverridecommandlockouts
\makeatletter
\def\ps@headings{%
	\def\@oddhead{\mbox{}\scriptsize\rightmark \hfil \thepage}%
	\def\@evenhead{\scriptsize\thepage \hfil \leftmark\mbox{}}%
	\def\@oddfoot{}%
	\def\@evenfoot{}}
\makeatother
\usepackage{nicefrac}
\usepackage{xr}
\usepackage{verbatim}
\usepackage{placeins}
\usepackage{soul}
\usepackage[para]{threeparttable}
\usepackage{graphicx}
\usepackage{algorithm}
\usepackage[noend]{algpseudocode}

\usepackage{subcaption}
\usepackage{url}
\usepackage{graphicx}

\usepackage{epsfig}
\usepackage{url}
\usepackage{stfloats}
\usepackage{enumerate}
\usepackage{longtable}

\usepackage{latexsym}
\usepackage{verbatim}
\usepackage{amssymb}

\usepackage{xspace}

\usepackage{xcolor}

\usepackage{verbatim}
\usepackage{caption}
\usepackage{subcaption}
\usepackage{amsmath}
\usepackage{amssymb}
\usepackage{wrapfig}
\usepackage{tabularx}
\usepackage{lscape}
\usepackage{float}
\usepackage[utf8]{inputenc}

\usepackage[utf8]{luainputenc}
\usepackage[T1]{fontenc}
\usepackage{pgfgantt}
\usepackage{footnote}
\usepackage{algpseudocode}
\usepackage{algorithm}
\usepackage{multicol}
\usepackage{multirow}
\usepackage{tablefootnote}
\usepackage{mathtools}
\usepackage{graphicx}
\usepackage{caption}
\usepackage{subcaption}
\usepackage{amsmath}
\usepackage{tikz}
\usepackage{float}
\usepackage{bm}
\usepackage{url}
\usepackage{amsthm}

\newcommand{\ignore}[1]{}
\usetikzlibrary{matrix,shapes,arrows,positioning,chains, calc}

\newcommand{\specialcell}[2][c]{%
	\begin{tabular}[#1]{@{}c@{}}#2\end{tabular}}

\mathchardef\mhyphen="2D

\usepackage{pbox}
\newtheorem{definition}{Definition}

\mathchardef\mhyphen="2D
\newcommand{\sk}{\ensuremath {\mathit{sk}}{\xspace}}
\newcommand{\pk}{\ensuremath {\mathit{PK}}{\xspace}}
\newcommand{\as}{\ensuremath {\leftarrow}{\xspace}}

\mathchardef\mhyphen="2D

\newcommand{\onlykg}{\ensuremath {\mathsf{Kg}}{\xspace}}
\newcommand{\onlysig}{\ensuremath {\mathsf{Sig}}{\xspace}}
\newcommand{\onlyver}{\ensuremath {\mathsf{Ver}}{\xspace}}

\newcommand{\SGNCOREKG}{\ensuremath {\mathsf{SGN.Kg}}{\xspace}}

\newcommand{\SGNCORESIG}{\ensuremath {\mathsf{SGN.Sig}}{\xspace}}
\newcommand{\SGNCOREVER}{\ensuremath {\mathsf{SGN.Ver}}{\xspace}}

\newcommand{\SGNONLY}{\ensuremath {\mathsf{SGN}}{\xspace}}

\newcommand{\flatss}{\ensuremath {\mathsf{FROG}^*}{\xspace}}
\newcommand{\flatssSig}{\ensuremath {\flatss\mathsf{.Sig}}{\xspace}}
\newcommand{\flatssVer}{\ensuremath {\flatss\mathsf{.Ver}}{\xspace}}
\newcommand{\flatssKg}{\ensuremath {\flatss\mathsf{.Kg}}{\xspace}}

\newcommand{\flats}{\ensuremath {\mathsf{FROG}}{\xspace}}
\newcommand{\flatsSig}{\ensuremath {\mathsf{FROG.Sig}}{\xspace}}
\newcommand{\flatsVer}{\ensuremath {\mathsf{FROG.Ver}}{\xspace}}
\newcommand{\flatsKg}{\ensuremath {\mathsf{FROG.Kg}}{\xspace}}
\newcommand{\LWEOTS}{\ensuremath {\mathsf{\SGNONLY}}{\xspace}}

\newcommand{\mmm}{\ensuremath {\mathsf{MMM}}{\xspace}}

{\xspace}

\newcommand{\eat}[1]{}                % Comment out large swaths of text
\newcounter{linecounter}

\usepackage[hang,flushmargin]{footmisc}

\usepackage{pifont}% http://ctan.org/pkg/pifont

\usepackage{algpseudocode}% http://ctan.org/pkg/algorithmicx

\DeclareUnicodeCharacter{202C}{}

\begin{document}

%\title{Lightweight Digital Signatures for IoT Systems}

\title{\flats:~{Forward-Secure Post-Quantum Signature}}
\addtolength{\topmargin}{0.05in}
\author{
	\IEEEauthorblockN{Attila A. Yavuz}
\IEEEauthorblockA{University of South Florida\\
	Tampa, Florida, USA \\
	attilaayavuz@usf.edu}
\and
	\IEEEauthorblockN{Rouzbeh Behnia}
\IEEEauthorblockA{University of South Florida\\
	Sarasota, Florida, USA \\
	behnia@usf.edu}}
\maketitle

\newcommand\blfootnote[1]{%
	\begingroup
	\renewcommand\thefootnote{}\footnote{#1}%
	\addtocounter{footnote}{-1}%
	\endgroup
}

\begin{abstract}
Forward-secure signatures guarantee that the signatures generated before the compromise of private key remain secure, and therefore offer an enhanced compromise-resiliency for real-life applications such as digital forensics, audit logs, and financial systems. However, the vast majority of state-of-the-art forward-secure signatures rely on conventional intractability assumptions and therefore are not secure against quantum computers. Hash-based signatures (HBS) (e.g., XMSS) can offer forward-secure post-quantum security.  However, they are efficient only for a pre-defined number of messages to be signed and incur high key generation overhead, highly expensive signing, and large signature sizes for an increasing number of messages. It is an open-problem to develop quantum-safe forward-secure signatures that are efficient and practical with a signing capability scalable to their security parameters. 

In this work, we propose a new series of post-quantum signatures that we call \flats~({\em Forward-secuRe pOst-quantum siGnature}). Unlike HBS alternatives, \flats~can achieve  highly computational efficient signatures with sub-linear key/signature sizes and (practically) unbounded signing capability. This is achieved by transforming suitable post-quantum signatures into forward-secure setting via MMM~generic constructions.  Specifically, we investigated the transformation of prominent post-quantum secure signatures such as Dilithium, WOTS and BLISS with MMM variants. Our experiments indicate that \flats~outperforms XMSS for the vast majority (if not all for large number of messages) performance metrics. We also discuss one-time variants of  these base signature schemes that can push the performance of \flats~to the edge.  Overall,  \flats~shows a better performance than the existing alternatives with forward-security, and therefore is an ideal alternative for the standardization efforts for forward-secure post-quantum signatures.
\end{abstract}

%\blfootnote{*Work done when the third author was employexd at University of South Florida.}
%{\footnotesize* The first and second authors contributed equally to this work.}

\section{Introduction}\label{sec:Introduction}%
%\vspace{-1mm}
Forward-security ensures that the past messages are protected even if the current secret key is exposed.  Forward-secure digital signature schemes~(e.g., \cite{FWSigBasis_Anderson97}) are key-evolving signatures that periodically update the private key and delete the previous key. This prevents an adversary who compromises the system from forging the previously computed digital signatures on past messages, since their corresponding keys were erased. Forward-secure digital signatures enhance the breach and compromise-resiliency of digital signatures. Hence, they play a critical role in many real-life applications such as secure audit logging, digital forensics, financial applications, public key certificate infrastructures, and many others. 

\subsection{The State of the Art and Its Limitations} \label{subsec:}
Several forward-secure digital signatures have been proposed with different performance trade-offs (e.g.,  \cite{FWSig_BellareMiner99,FWSecItkis_Reyzin_01,FWSecKozlov_Reyzin_02}). There is generally a trade-off between signature and public key sizes, update frequency and computational efficiency, and number of items that can be signed for some variants. Forward-secure signatures generally are computationally costly compared to standard signatures, and have larger signature and/or key sizes. Some of the forward-secure signatures offer extended properties such as aggregation~\cite{BAFYavuzNing09,Yavuz:2012:TISSEC:FIBAF,FssAggNew,FAS_Asymptotic,Yavuz:CORE:2020}, untrusted update~\cite{FWSecBoyenUntrusedupdate:2006,FWSigUntrustedUpt_CCS07} ,   selective verification~\cite{Logging_VerifiableExcerpt}, and group signatures~\cite{FWSEc:LatticeGrou:2019}. These features might increase the efficiency of the schemes in  one front, while incurring other cost and constraints in an another  (e.g., more compact signatures but larger public keys and/or more computational overhead). There are also generic forward-security frameworks (e.g.,~\cite{FWSigfromAnyScheme_Hugo2000,MaMiMi02}) that can transform any standard signature into a  forward-secure signature. However, these generic transformations might lead inefficient results (e.g.~\cite{FWSigfromAnyScheme_Hugo2000}), and specially tailored constructions might outperform them with careful designs (e.g.~\cite{FWSec:TightReduction:2020}) for some performance metrics. 

The vast majority of  the existing forward secure signatures rely on conventional intractability problems (e.g., factorization, discrete logarithm problem (DLP), elliptic curve (ECDLP)). However, it is well-known that the conventional signature schemes are vulnerable to the emerging quantum computers, and therefore, NIST has initiated the second round of standardizations for post-quantum    cryptography\footnote{\url{https://csrc.nist.gov/projects/post-quantum-cryptography/}}. The existing post-quantum digital signatures, compared to   their conventional counterparts, are known to be   costly in terms of communication, computation and energy consumption. Forward secure version of such constructions are expected to be  even costlier than the original schemes. 

The forward-security for post-quantum signatures have been mainly investigated for hash-based digital signatures (HBS) ~\cite{OTS_Lamport_79}. To our best knowledge, the other NIST standardization   candidates have not yet been explored extensively from the forward-security perspective. For instance, it is clear that a generic forward-secure transformation (e.g., ~\cite{FWSigfromAnyScheme_Hugo2000,MaMiMi02}) for  many of these candidates yield highly inefficient constructions due to their very large signature and/or public key sizes.  HBS (e.g., Lamporr~~\cite{OTS_Lamport_79}) are among the first signature schemes with  a post-quantum security. There are  number of efficient one-time HBS (e.g., ~\cite{OTS_r_time_HORSED_2006,HORS_BetterthanBiBa02,OTS:WOTS,OTS:WOTSPlus:2013}) but they can only sign one or a few messages per key pair. Recently, a NIST candidate (SPHINCS~\cite{SPHINCS:Bernstein:2015}) addressed this limitation   with a stateless design, albeit suffering from   large signature sizes (e.g., 41 KB).  

Forward-secure signatures are by design stateful. Hence, to our knowledge, at this point, the HBSs such as  XMSS~\cite{XMSS:Buchmann:2011} and LMS~\cite{LMS:OriginalPatent,LMS:RFC}  are the most efficient forward-secure post-quantum secure alternatives. These schemes are usually efficient only for small to moderate number of messages to be signed. However, once the total number of messages to be signed  increases, the computational overhead, and especially the cost of key generation, grows significantly. Both of these schemes rely on hierarchical variants (e.g., HSS-LMS) to handle signing capabilities over $t>2^{40}$. (we refer them as Hierarchical HBS (HHBS)).  In any cases, HHBSs generally require $t$ to be pre-defined, and their key/signature generation overhead and signature sizes grow extremely large  for a big $t  $ .  There is a need for post-quantum forward-secure signatures that are efficient, practical and can scale for large number of messages to be signed. 

\subsection{Objectives and Desirable Properties}
The goal of this work is to create  a series of  post-quantum forward-secure signatures that achieve highly efficient key and signature generation while maintaining sub-linear signature, private key and constant public key sizes. We aim that our schemes should be at least competitive or outperform existing alternatives for small/moderate $t$ values (for some performance metrics), while significantly outperform them when $ t $ grows   for the   all performance metrics.  We outline some of the desirable properties of  \flats. 

\begin{itemize}
	
	\item ~{\em \underline{Efficient Key Generation }}: For $ t=2^{64} $, the key generation of \flats, instantiated with  the lattice-based signature in \cite{dilithium}~is almost $ 2200\times $ faster than the most efficient key generation instance of XMSS-MT variant for key generation (with 12 layers) with the maximum of  $ 2^{60} $ signature generation capability.   This difference significantly grows for XMSS-MT variants with lower number of sub-trees  (e.g., 3 or 6). 
	
	\item ~{\em \underline{Efficient Signature Generation }}: The signature generation of \flats~is two signature generation  plus three key generation overhead of the underlying signature scheme along with a constant number of hash calls. This cost is again independent from both the total number of messages to be signed or number of messages signed so far, making \flats~signing magnitudes of times more efficient than XMSS-MT. For example, for $t=2^{64}$, the lattice-based instantiation of \flats~based on \cite{dilithium} (with AVX optimization) is $  \approx 15 \times $ faster than most efficient  XMSS-MT  variant for signature generation (with 12 layers).  The hash-based instantiation of \flats~based on \cite{WOTSP} is $  \approx 8 \times $ faster than the most efficient XMSS-MT  variant (with 12 layers). 
	
	\item ~{\em \underline{Efficient Signature Verification}}:  The signature verification of \flats (with basic MMM instantiation)~is two signature verification  plus   a constant number of hash calls.  For example, for $t=2^{64 }$, the lattice-based instantiation of \flats~based on \cite{dilithium} is $  \approx 17\times $ faster than most efficient  XMSS-MT  variant for verification (with 3 layers).  The hash-based instantiation of \flats~based on \cite{WOTSP} is $  \approx 8 \times $ faster than the most efficient XMSS-MT  variant (with 3 layers). 
	
	\item ~{\em \underline{Compact Public Key}}: The public key of \flats~is just a hash value ( half size of XMSS-MT variants), and therefore is optimal.
	
	\item ~{\em \underline{Competitive Signature Sizes}}:  The signature sizes in  some \flats~variants (e.g., hash-based) could slightly larger than those variants in XMSS-MT with lower number of layers (e.g. $ t=2^{60} $ with three layers), however, usually such variants of XMSS-MT have a very costly key generation  and signature generation algorithms.  For example, for instantiation of \flats~with Dilithium \cite{dilithium}, the signature size is $ 13,624 $  bytes where in XMSS-MT with $ 12 $ layers, signature size is $ 27,688 $ bytes. The signature size becomes more favorable for XMSS-MT with lower layers. We investigated another variant of \flats, called \flatss (iterative product composition of MMM)~that enjoys shorter signatures and private keys but with a more expensive key generation and signature verification.  Remark that in many cases, \flatss~is still significantly more computationally efficient than XMSS-MT variants. 
\end{itemize}

\noindent\textbf{Limitations:} As aforementioned, our constructions (\flats~and \flatss) outperform XMSS-MT variants in almost all the metrics. However, our schemes suffer from larger private key sizes. For instance, for the hash-based instantiations of \flats, the private key size could be as large as $ 715,840 $ bytes, which is about $ 18\times$ larger than the largest private key size for the instantiated XMSS-MT variants. \flatss  has a smaller private sizes but with more computational overhead. However, we think that this is a very favorable trade-off since unlike public key size, the private key size does not incur any (online) communication overhead.

\section{Preliminaries}\label{sec:prelim}
%For our lattice-based signature scheme, we work on a ring  $R = \Z_q[x]/f(x)$ where $ f(x)= x^n+1 $. 
Operators $||$ and $|x|{=\log x}$ denote the concatenation and the bit length of variable $x$, respectively and  $ \log $ denotes $ \log_2 $.
\begin{definition}\label{def:Generic} \normalfont 
	
	A  signature scheme consists of three algorithms $  \SGNONLY = ({\onlykg,\onlysig,\onlyver}) $ defined as follows.
	
	\begin{itemize}\normalfont
		\item[--] $ (\sk, \pk) \leftarrow \SGNCOREKG(1^\kappa) $: Given the security parameter $ \kappa $, it outputs the private and public key pair $ (\sk,\pk) $.
		
		\item[--] $ \sigma \leftarrow \SGNCORESIG(m,\sk) $:
		Given the message   $ m $ and the signer's private key  $\sk$, it outputs the signature $ \sigma $.
		\item[--] $ b \leftarrow \SGNCOREVER(m, \sigma, \pk) $: Given a message-signature pair   ($m,\sigma $), and public key  $\pk$, outputs $b \as \{0,1\}$
		%if the signature is verified it outputs $b=1$ else $b=0$. %it outputs a decision bit  $b \as \{0,1\} $.
	\end{itemize}
\end{definition}
In ordinary digital signatures, if the private key of the signer is compromised, all the signatures that are generated by the signer (past and future) become forgeable.  In forward secure signatures, the previously computed signatures remain unforgeable even if the current private key of the signer is compromised. 
 
 \subsection{Malkin, Micciancio and Miner (MMM) Scheme}
 \mmm~\cite{MaMiMi02} is a series of constructions to transform standard digital signatures $\SGNONLY$ to forward secure   signature.  \mmm~is composed of the sum (referred to as $\bigoplus$) and product (referred to as $\bigotimes$) composition algorithms. In the sum composition, given a digital signature $\SGNONLY$ with $t$ forward-secure signing capability, the sum composition can generate $2t$ signatures in total. For instance, given a standard one-time signature, by iteratively applying sum composition  $\log t$ times, one can obtain a forward-secure signature with $t$ signing capability.  In the product composition, given a digital signature $\SGNONLY$ with $t$ forward-secure signing capability, the product composition can generate $t^{2}$ signatures in total.

The main  \mmm~construction has an upper tree and multiple lower trees, which are created with the iterative execution of sum composition and are linked with the upper tree with the product composition. Lower trees are generated on the go with increasing levels of height.   This minimizes the key generation/update cost and makes the overhead of \mmm~depend on "messages signed so far" but not the total number of time periods available. Thus, the lower trees are created as needed on the go. \mmm~increases the levels in each iteration to sign more messages with a slight increase in cost (generation of a higher tree). Overall, in \mmm, the costs are either logarithmic or constant with respect to $t<T$ (i.e., $t$ is the total number of messages signed so far, and $T$ is the maximum number of signatures) and the size of the $\pk$ is a small-constant. We refer curious reader to \cite{MaMiMi02} for the details.

Another interesting composition in  \cite{MaMiMi02} is the iterated product composition.  In this construction, a new forward secure scheme is achieved by applying the sum composition once and then iterate the product composition to this two-time signature to get $ 2^{2^{\log\log t}} $ and get $ t $  time signature scheme.

\section{The Proposed Scheme FROG}  \label{sec:contrib}
We   instantiate  the generic constructions given in \cite{MaMiMi02} with the existing efficient lattice-based  \cite{dilithium,BLISS} and hash-based \cite{WOTSP}  post-quantum signatures to create a    series new post-quantum forward-secure signature we call \flats~({\em Forward-secuRe pOst-quantum siGnature}) . 

%\noindent \textbf{Design Rationale}: 

\subsection{Design Rationale} 
In our design, we aim to achieve high computation efficiency while keeping  sub-linear signature/key sizes, all with signing capabilities scalable to the security parameter (e.g.,  $t = 2^{\kappa}$).  However, following NIST guidelines, we set $ t=2^{64} $ for \flats. 

(i) The sub-linear key sizes in forward-secure schemes are generally achieved by a tree structure. For instance, in XMSS-MT, given a pre-determined (fixed) $t=2^{h}$  ($ h  $ is the height of the tree) number of messages to be signed, $d$ XMSS  sub-trees are computed during key generation, each of height $l$, such that $h=l\cdot d$. A Merkle-tree is constructed on top these sub-trees to compute the final public key that can verify the generated signatures, which include the authentication path from leaves to the root. Note that the tree structure in XMSS-MT is rigid and $t$ is pre-determined. The signature sizes, signature generation and especially key generation overhead grow substantially for larger $t$ values, and if it reaches $t$, the system must be re-initialized.  HSS-LMS~also follows a similar strategy.
In our design, we depart from HHBS approaches, but instead rely on \mmm~\cite{MaMiMi02} that generates hierarchical trees {\em as needed and on-the-fly}. In particular, we   harness the sum composition  iteratively and in conjunction with product composition to obtain practical performance results. However, as mentioned before,  \mmm~generic framework might lead to  inefficient forward-secure schemes if the underlying primitive and composition methods are not selected carefully. 

(ii) We observe that {\em on-the-fly} tree construction requires generating a  fresh private/public key pair per update, and therefore, the key generation overhead must be minimized.  That is, the signing overhead of on-the-fly constructions depend on the key generation  of the underlying signature primitive as much  its signing overhead. We identified that Dilithium~\cite{dilithium} and BLISS~\cite{BLISS} offer efficient key generation and signing, making them appropriate choices for the iterated sum and product compositions. For the sake of providing a more leveled comparison with HHBS schemes, we also instantiated schemes based on a relatively recent one-time hash-based signature in \cite{WOTSP}. In iterated sum and product compositions, we create a new private/key pair for each item to be used, and then another two pairs for the future tree elements in an amortized manner.  

(iii) New public keys must be relayed to the verifier for each update, and therefore both the signature and public key sizes must be minimized. We selected BLISS \cite{BLISS}  due to its parameters sizes which offers a more compact size than the existing post-quantum signature alternatives.  Our approach also shed a light of how existing NIST post-quantum signature candidates perform under some generic forward-secure transformations. Our observation indicates that only the candidates whose total signature plus public key size is relatively small might yield efficient results. In this regard, the lattice-based candidates such as Dilithium~\cite{ducas2018crystals} and Falcon~\cite{fouque2018falcon} seem only alternatives. We will later discuss our future plan to devise a one-time lattice based signature similar to \cite{DBLP:journals/joc/LyubashevskyM18} and with the optimizations proposed in \cite{dilithium} and the potential performance gains from such construction. 

%\textbf{The Description of  Main Instantiation}:  
 \subsection{The Description of  Main Instantiation}
\flats~is constructed by performing an iterative sum and product composition of \SGNONLY. Let  $\bigoplus$ and $\bigotimes$ denote the application of sum and product composition defined in Section \ref{sec:prelim} only once. We denote composing two instantiations of \SGNONLY~with the sum composition as $\SGNONLY \bigoplus \SGNONLY$. Similarly,  $\SGNONLY \bigotimes \SGNONLY$ denotes two instantiations of a signature scheme with the sum composition. $\SGNONLY^{\bigoplus}_{K}$ means an iterative composition of \SGNONLY ~with itself $K$ times. Recall that $1\leq t \leq 2^{\kappa}-1$ denotes the total number of signatures signed {\em so far}.  For example, in XMSS-MT, $1\leq t \leq K << 2^{\kappa}-1$, whereas in \flats~$t$ can approach to $2^{\kappa}$.  We can express the \flats~as follows:

\begin{equation*}
\SGNONLY_{\log t}^{\bigoplus} \bigotimes \{ \SGNONLY_{i}^{\bigoplus} \}_{i=0}^{t}
\end{equation*}

\mmm~has an {\em } upper-tree and multiple lower-trees, which are constructed on-the-fly as needed:

 (i) We first create a forward-secure signature scheme with $\log t$ signing capability by iterating $\SGNONLY_{\log t}^{\bigoplus}$ via sum composition. These private/public key pairs are the leaves of {\em the upper-tree}, and are used to certify the public keys of the signatures that will be used to verify the data items in the lower-trees. (ii) There will be multiple lower-trees that will be created as new data items to be signed arrive. Each lower-tree $1\leq i \leq t$ will be a new forward-secure iterative sum composition $ \LWEOTS_{i}^{\bigoplus}$ that can sign $2^{i}$ data items. (iii) The leaves of upper-tree and the lower-tree are connected with a product composition. That is, once the leaves (i.e., key pairs) of $ \LWEOTS_{i-1}^{\bigoplus}$ are depleted, we create a new lower-tree $ \LWEOTS_{i}^{\bigoplus} $ and compose it with the corresponding leave of the upper-tree by calling the product composition. (iii) We adopt the amortized update strategy to ensure that transition from one lower-tree to the next one does not incur heavy key generation overhead.  In each update operation for  $ \LWEOTS_{i}^{\bigoplus} $, we also generate two key pairs (the leaves) for the next tree $i+1$. Hence, once the key pairs for $i$'th lower-tree are depleted, the keys for new tree will be ready, and heavy (bulk) key generation is not needed. This approach enables a uniform update efficiency and is essential to handle large lower-tree sizes. However, it requires storing initial keys for the next lower-tree, and therefore  increases the private key size. This overhead only grows with $t$  as $O(\kappa \cdot \log t^{2})$, and therefore is space-efficient. For small-number of messages to be signed, one can omit amortization and perform batch update only for new lower-trees to be generated. 

%\textbf{ Alternative Instantiations}: 
\subsection{Alternative Instantiations}
(i) As aforementioned, one can use different signature schemes (either one-time or polynomially bounded)  to be used in the construction. However, careful consideration needs to be taken in order to ensure that the resulting forward-secure scheme will be efficient.   For instance, one can imagine the use of W-OTS variants~\cite{OTS:WOTS,OTS:WOTSPlus:2013}  to offer different performance trade-offs. In W-OTS, the length of hash-chain $w$ poses a trade-off between the signature/public key sizes and computational overhead. We have investigated different $w$ values and signature/public key sizes for W-OTS instantiated with \mmm. In order to have comparable signature sizes with \flats, $w$ value must be relatively large (e.g. $  > $400), and this makes the key generation and signing of W-OTS rather expensive~for $\kappa=100-110$. It might be possible that W-OTS variants offer desirable trade-offs for different $\kappa$ and some applications might prefer a hash-based signature as a base scheme as opposed to a lattice-based constructs. Hence, we capture the possibility of any future OTS that has smaller signature and/or  public key sizes that may rely on a different building blocks.

%
%(i) One can use a different OTS instead of \LWEOTS , whose advantages for our particular constructions have been explained in our design rationale. Yet, for instance, one can imagine the use of W-OTS variants~\cite{OTS:WOTS,OTS:WOTSPlus:2013} instead  of \LWEOTS~to offer different performance trade-offs. In W-OTS, the length of hash-chain $w$ poses a trade-off between the signature/public key sizes and computational overhead. We have investigated different $w$ values and signature/public key sizes for W-OTS instantiated with \mmm. In order to have comparable signature sizes with \flats, $w$ value must be relatively large (e.g., >400), and this makes the key generation and signing of W-OTS more expensive than that of \LWEOTS~for $\kappa~100-110$. It might be possible that W-OTS variants offer desirable trade-offs for different $\kappa$ and some applications might prefer a hash-based signature as a base scheme as opposed to a lattice-based constructs. Hence, we do not exclude the use of  hash-based OTS with iterative sum and product compositions as a viable alternative. We also capture the possibility of any future OTS that has smaller signature and/or  public key sizes that may rely on a different building blocks.

(ii-iii) We consider hybrid approaches, wherein the upper tree is replaced by a (H)HBS and the lower-tree is generated with iterative constructions as described in the main scheme. This may lead an efficient constructions as (H)HBS (e.g., LMS~\cite{LMS:RFC}) is efficient for small-fixed number messages to be signed, and the upper-tree harbors $\kappa$ key pairs at most. Another alternative is to compute a hash-chain for any selected signature, and place $\kappa$ public keys as a part of the master public key. This approach makes the master public key large, but in return, it makes \mmm~signature size much smaller as the hash chain elements replace the leaves of upper-tree, and only the signature is transmitted but not the public keys. 

\begin{table*}[t!]
	\centering
	\vspace{ 2mm}

	\caption{Analytical performance of \flats~and \flatss} \label{tab:analytical}
	%	\vspace{-2mm}
	\begin{threeparttable}
		\begin{tabular}{| c || c | }
			\hline
			
			\textbf{Operation/Parameter} & \specialcell[]{\textbf{Computation/Storage Cost}}\\ \hline \hline 
			\multicolumn{2}{|c|}{\textbf{MMM}} \\ \hline\hline 
			$ \flatsKg  $ 	& $ 2 \; \SGNCOREKG $ \\ \hline
			$ \flatsSig $ 	& $   3\;  \SGNCOREKG+2\;  \SGNCORESIG+2H  $ \\ \hline
			$ \flatsVer $  	& $  2\; \SGNCOREVER+(\log  \kappa + \log t )H  $ \\ \hline
			$\flats_{\sigma}$& $ 2|\sigma|+4|pk|+( \log \kappa +\log t +1)\cdot |H|$\\ \hline 
			$\flats_{pk}$ & $  |H|$\\ \hline 
			$\flats_{sk}$ & $  (2+ \log \kappa)\cdot |sk|+6|pk|+4\cdot  \log \kappa \cdot |H|  + 3\log t \cdot |H| + \kappa (\log t)^{2}$\\ \hline \hline 
			\multicolumn{2}{|c|}{ \textbf{Iterated Product  Composition}} \\ \hline\hline 
			$ \flatssKg  $ 	& $\log \log t (5\; \SGNCOREKG+\SGNCORESIG+2 \kappa \cdot H)  $ \\ \hline
			$ \flatssSig $ 	& $  3 \; \SGNCOREKG+2 \; \SGNCORESIG+ (\kappa +1) \cdot H $ \\ \hline
			$ \flatssVer $  	& $ 2 (\log \log t \cdot \SGNCOREVER+H)   $ \\ \hline
			$\flatss_{\sigma}$& $ 2|\sigma|+4|pk|+\kappa $\\ \hline 
			$\flatss_{pk}$& $  |H|$\\ \hline 
			$\flatss_{sk}$& $ \log \log t (2 |sk| + 6|pk|+4\kappa)      +       \kappa \cdot (\log  t)^{2}+|sk|\cdot \log \kappa$\\ \hline 
			
		\end{tabular}
		\begin{tablenotes}[flushleft]\scriptsize{  % switch locally to single-spacing
				%				\item $\ddagger$ The sizes are in terms of \textbf{Bytes}, if otherwise not specified.
%				
%				\item $ \dagger $ System wide parameters (e.g., p,q,$\alpha$) for each scheme are included in their corresponding codes, and private key size denote to specific private key size. 
				
				%				\item The cost of hash-based schemes are estimated based on the cost of a single hash operation.
				
				%				\item The parameters for multiple-time signatures are selected such that they can sign XXX messages with a single key pair.
			}
		\end{tablenotes}
	\end{threeparttable}
	% }
	%	\vspace{-3mm}
\end{table*}

(iv) All of the above techniques receive significant computational benefit from pre-computation methods with an expense of  larger memory usage. For example, in amortized update strategy, the generated keys are independent from the messages, and therefore can be pre-computed and stored to be used later. This accelerates signature generation as a small and constant number of key update operations are performed beforehand. Parallel computing (e.g., GPUs, FPGAs) can be used to accelerate the batch generation of pre-computed elements and then fed into online  computations. 

\subsection{Security Argument from Base Scheme OTS and MMM}
\flats~is a transformation of efficient post-quantum one-time signatures into multiple-time forward-secure signatures via MMM~\cite{MaMiMi02}. Hence, the security of \flats~directly follows from the sum and/or product composition proofs of MMM~in~\cite{MaMiMi02} provided that the base OTS schemes are secure. We instantiate \flats~with well-proven OTSs as outlined, and therefore \flats~schemes are as secure as MMM with its corresponding base OTS schemes.

\begin{table*}[t!]
	\centering
	\vspace{ 2mm}

	\caption{Experimental performance comparison of  \flats, \flatss~and its counterparts on a commodity hardware} \label{tab:Laptop}
	%	\vspace{-2mm}
	\begin{threeparttable}
		\begin{tabular}{| c || c | c | c | c |  c | c | }
			\hline
			\textbf{Scheme} & \specialcell[]{\textbf{Key Generation}} &  {\textbf{Sign}} & {\textbf{Verify }} & \specialcell[]{\textbf{Signature}\\  \textbf{(B)}} & \specialcell[]{\textbf{Public Key} \\ \textbf{(B)}} & \specialcell[]{\textbf{Private Key} \\ \textbf{(B)}}\\ \hline \hline 
			
			%			RSA & 11374.27 & 384 & 384 & 275.65 & 386 & 11649.92 \\ \hline
			
			XMSS-MT-SHA2\textunderscore20/2\textunderscore256 & $9,236,557,672$ & $24,554,349$ & $5,186,460$ & $4,963$ & $64$ & $5,998$\\ \hline
			XMSS-MT-SHA2\textunderscore20/4\textunderscore256 & $729,631,517$ & $14,364,265$ & $10,188,082$ & $9,251$ & $64$ & $10,938$\\ \hline
			XMSS-MT-SHA2\textunderscore40/2\textunderscore256 & $9,404,925,498,412$ & $26,628,986$ & $5,377,454$ & $5,605 $& $64$ & $9,600$ \\ \hline 
			
			XMSS-MT-SHA2\textunderscore60/3\textunderscore256& $14,234,635,667,761$ & $29,584,259$ & $7,619,770$ & $8,392$ & $64$ & $16,629$ \\ \hline 
			
			%			XMSS~\cite{} & $2^{\kappa}$& & & & & & \\ \hline 
			
			XMSS-MT-SHA2\textunderscore60/6\textunderscore256& $31,682,214,982 $& $31,391,553$ & $16,521,985$ & $14,824$ & $64$ & $24,507$\\ \hline
			
			XMSS-MT-SHA2\textunderscore60/12\textunderscore256 &  $1,946,231,536$ &$ 15,474,825$ & $33,375,298$ & $27,688$ & $64$ & $38,095$ \\ \hline\hline 
			\multicolumn{7}{|c|}{\textbf{ MMM}} \\ \hline\hline
			\flats-BLISS&  $2,102,770$ & $12,517,153$ & $999,972$ & $7,054$ & $32$ & $80,076$ \\ \hline 
			\flats-Dilithium&  $815,322$ & $5,544,419$ & $994,438$ & $ 13,624$ & $ 32$ & $ 634,176$ \\ \hline

			\flats-Dilithium-AVX2 &  $261,832$ & $1,369,462$ & $432,882$ & $ 13,624$ & $ 32$ & $ 634,176$ \\   \hline 
			
			\flats-WOTS+(SHA256) &  $810,768$ & $2,031,100$ & $959,158$ & $ 27,872$ & $ 32$ & $ 715,840$ \\   \hline 
			\flats-WOTS+(SHAKE256) &  $2,223,760$ & $5,559,400$ & $2,223,760$ & $ 27,872$ & $ 32$ & $ 715,840$ \\   \hline \hline
			\multicolumn{7}{|c|}{{\textbf{Iterated Product  Composition}}} \\ \hline\hline
			\flatss-BLISS&  $62,828,244$ & $12,782,583 $ & $5,113,672 $ & $4,766$ & $32$ & $10,640$ \\ \hline 
			\flatss-Dilithium&  $ 28,642,638$ & $5,830,749$ & $ 5,080,468$ & $ 11,305$ & $ 32$ & $ 85,362$ \\ \hline

			\flatss-Dilithium-AVX2 &  $10,306,122$ & $1,655,792$ & $1,711,132$ & $ 11,305$ & $ 32$ & $ 85,362$ \\   \hline 
			
			\flatss-WOTS+(SHA256) &  $ 17,804,064$& $ 2,296,530$ & $ 4,868,788$ & $ 25,552$ & $ 32$ & $ 107,040$ \\   \hline 
			\flatss-WOTS+(SHAKE256) &  $ 40,027,680$ & $5,829,010$ & $13,346,740$ & $ 25,552$ & $  32$ & $ 107,040$ \\   \hline

			%			\multirow{3}{*}{\eda} & 1 & \multirow{3}{*}{\textbf{2 425}} & \multirow{3}{*}{\textbf{32}} & \multirow{3}{*}{\textbf{32}} & \multirow{3}{*}{52 872} & 96 & \multirow{3}{*}{\textbf{55 297}} \\ 
			%			
			%			& $2^{10}$&  &  & & & 64KB & \\ 
			%			
			%			& $2^{17}$&  &  & & & 8MB & \\ \hline
			
		\end{tabular}
		\begin{tablenotes}[flushleft]\scriptsize{  % switch locally to single-spacing
				%				\item $\ddagger$ The sizes are in terms of \textbf{Bytes}, if otherwise not specified.
				
 		\item All the above schemes, except for the instantiations of Dilithium, which gives138-bits of security, provide 128-bit of security.  
% 		 System wide parameters (e.g., p,q,$\alpha$) for each scheme are included in their corresponding codes, and private key size denote to specific private key size.
				
				%				\item The cost of hash-based schemes are estimated based on the cost of a single hash operation.
				
				%				\item The parameters for multiple-time signatures are selected such that they can sign XXX messages with a single key pair.
			}
		\end{tablenotes}
	\end{threeparttable}
	% }
	%	\vspace{-3mm}
\end{table*}
\section{Performance Analysis} \label{sec:PerformanceAnalysis}
In this section, we compare the performance of  different instantiations of \flats~with  its counterparts. We also discuss the performance of some of the selected alternative constructions described in Section \ref{sec:contrib}. As mentioned, we also instantiated forward secure schemes using the iterated product composition technique presented in \cite{MaMiMi02}. For the sake of clarity we denote instantions with this technique as \flatss.

%Private/public key sizes, signature size and signature generation/verification costs of \core~and its counterparts

%We now show performance analysis on $\core$ and it's counterparts.

%\subsection{Parameter Selection}
%Our $\core$ scheme and it's counterparts have $\kappa$ as 128-bit security. We selected our $K$ as $2^{10}$ (we believe it is a good balance for efficiency and verification). 
%Our $|q|$ is 256-bits. m is selected to be 260 for Zaeverucha et al. For HORS and HORSE, t is 1024 and $u$ is 24. We utilized the suggested/base parameters for the SPHINCS and XMSS schemes  

\subsection{Analytical Performance}
We denote the signature, public key, and private key sizes of the base signature scheme as $|\sigma|$, $|pk|$ and $|sk|$, respectively. We denote the size of hash output as $|H|$.   $\flats_{\sigma}$, $\flats_{sk}, \flats_{pk}$ denote the signature, public key, and private key sizes of \flats.  $H$ denotes a hash operation (also a PRF call for the simplicity). \flatsSig~includes the amortized update cost for a given composition.  Following the guidelines given by NIST\footnote{https://csrc.nist.gov/Projects/Post-Quantum-Cryptography/faqs}, we set $ t=2^{64} $.  We present the analytical performance of our instantiations (\flats~and \flatss) in Table \ref{tab:analytical}.  As depicted, \flatss~provides compact signature and private key sizes with the cost of  an increased computational overhead.

\subsection{Experimental Performance Evaluation and Comparisons} \label{subsec:PerformanceEval}
We now elaborate the detail of our performance analysis and comparison with XMSS-MT. We use the similar instantiations of XMSS-MT in  \cite{XMSS:LMS:Compare:Panos:2017:IACR}. We note that as also highlighted in  \cite{XMSS:LMS:Compare:Panos:2017:IACR}, the HSS-LMS is slightly more computationally efficient than XMSS-MT, but we currently focus on XMSS-MT standard due to  its ease of test.  We compare \flats~and \flatss~and its counterpart in terms of (i) private key, signature and public key sizes, and (ii) key generation, signature generation,  and  signature verification.  We run our experiments on an i7 Kaby Lake equipped with a  2.9 GHz Quad-Core Intel Core and 8 GB RAM. 

We instantiated both  \flats~and \flatss with three signature schemes: (i) Dilithium \cite{dilithium}: For this scheme we used the recommended parameters set  (see Table 2 in  \cite{dilithium}) which provides 138-bits of security with signature size and public key size  of 2701  and 1472 bytes, respectively. We used this scheme as it is one of the prominent lattice-based signature schemes for NIST post-quantum standardization process. We use the reference implementation and the AVX2 implementation  (increases parallelism and throughput in floating point SIMD calculations). (ii) BLISS \cite{BLISS}: For this scheme we used BLISS-II parameters set (see Table 1 in \cite{BLISS}) which provides 128-bits of security. We note that we have selected BLISS due to its smaller signature size. The signature and public key sizes in our selected variants are $ 625 $ and 875 bytes, respectively. BLISS using Gaussian Sampling to produce a one-time masking term in the signature generation and therefore, obtains a better signature size as compared to Dilithium \cite{dilithium}, with a cost of signature generation performance. Additionally we note that the Gaussian Sampling  technique is susceptible to side-channel attacks. We stress that each BLISS private/public key pair is used only once (minimal side-channel attack risk). We used the reference implementation for BLISS.  (iii) WOTS$ ^+  $ \cite{WOTSP}: We used  a variant with 128-bits of security with $ w=4 $ and $ m=n=256 $.  We instantiated it with both  SHA256   and SHAKE256.

As depicted in Table \ref{tab:Laptop}, we compared our instantiations with six different instantiations of   XMSS-MT-SHA256    with different parameters sizes, since, based on our initial benchmark, XMSS-MT with  SHA256 showed significantly better performance than the one instantiated with SHAKE256.  We adopted the implementation from \cite{XMSSGit}. For XMSS-MT we considered $ t =2^{20}, 2^{40}, 2^{60} $. For $ t =2^{20} $, we considered  the number of subtrees to be $ 2 $ and $ 4 $.  For $ t =2^{40} $, we considered  the number of subtrees to be $ 2 $.  For $ t =2^{60} $, we considered  the number of subtrees to be $ 3 $ $ 6 $, and $ 12$.

Following NIST's recommendation,  both \flats~and \flatss~allow for up to $ t= 2^{64} $ signature generation. \flats~performs significantly better than all of XMSS-MT variants for key generation, signature generation and signature verification. However, as a trade-off, it has larger private key sizes and slightly larger signature sizes. However, since private key is only stored on the signer's machine, and does not affect communication overhead. Our \flatss~variants provides better signature and private key sizes, with the cost of added performance overhead on all other algorithms. however, it is still shown to be more efficient than all other XMSS-MT variants. 
 
 \section*{Acknowledgment}
The work of Attila A. Yavuz is supported by the NSF CAREER Award CNS-1917627 and an unrestricted gift via Cisco Research Award.

\bibliographystyle{IEEEtran}
\bibliography{crypto-etc,AttilaYavuz}

%\vspace{-1mm}
%\appendix \label{sec:appendix}%{Proof of Theorem~\ref{the:ULSSecurityTheorem}}
%\vspace{-1mm}
%\input{proof}

\end{document}